\documentclass[%
 reprint,
 superscriptaddress,
 showkeys,
 amsmath,amssymb,
 aps,
]{revtex4-2}

\usepackage{graphicx}
\usepackage{dcolumn}
\usepackage{bm}

\newcommand{\s}{\sigma}
\renewcommand{\l}{\ell}
\renewcommand{\d}{\delta}

\newcommand{\la}{\ell_{\rm act}}
\newcommand{\lb}{\ell_{\rm allo}}

\newcommand{\beq}{\begin{equation}}
\newcommand{\eeq}{\end{equation}}
\newcommand{\bea}{\begin{eqnarray}}
\newcommand{\eea}{\end{eqnarray}}

\begin{document}

\preprint{APS/123-QED}

\title{On the emergence of single versus multi-state allostery}

\author{Eric Rouviere}
    \affiliation{Center for Physics of Evolving Systems, Department of Biochemistry \& Molecular Biology, University of Chicago, Chicago, IL, USA, 60637}
    \affiliation{Graduate Program in Biophysical Sciences, University of Chicago, Chicago, IL, USA, 60637}
\author{Rama Ranganathan}%
\affiliation{Center for Physics of Evolving Systems, Department of Biochemistry \& Molecular Biology, University of Chicago, Chicago, IL, USA, 60637}
\affiliation{The Pritzker School for Molecular Engineering, University of Chicago, Chicago, IL, USA, 60637}

\author{Olivier Rivoire}
\affiliation{Center for Interdisciplinary Research in Biology (CIRB), Coll\`ege de France, CNRS, INSERM, Universit\'e PSL, Paris, France}

\begin{abstract}
Several physical mechanisms have been proposed to explain allostery in proteins. They differ by the number of internal states that they assume a protein to occupy, leaving open the question of what controls the emergence of these distinct physical forms of allostery. Here, we analyze a simplified model of protein allostery under a range of physical and evolutionary constraints. We find that two archetypal mechanisms can emerge through evolution: a single-state mechanism where ligand binding induces a displacement along a soft normal mode or a multi-state mechanism where ligand binding induces a switch across an energy barrier to a different stable state. Importantly, whenever the two mechanisms are possible, the multi-state mechanism confers a stronger allosteric effect and thus a selective advantage. This work defines the essential constraints on single or multi-state allostery, and sets the stage for a physical theory of its evolutionary origins.
\end{abstract}

\maketitle


Allostery, the change of activity of a macromolecule in response to a perturbation at a distance from its active site, is thought to be a ubiquitous feature of proteins. Initially described in the context of multimeric proteins~\cite{monod1963allosteric, monod1965nature,koshland1966comparison}, it is now understood to underlie the regulation of proteins with diverse structural architectures, from receptors to signaling proteins and metabolic enzymes~\cite{volkman2001two,kamata2004structural,may2007allosteric,changeux2005allosteric,helmstaedt2001allosteric}.

Efforts to explain how allostery works date back to decades ago with the phenomenological models of Monod, Wyman, Changeux (MWC model) and Koshland, Nemethy, Filmer (KNF model)~\cite{monod1965nature,koshland1966comparison}. These models postulate that each subunit of a multimeric allosteric protein occupies a small number of internal states between which transitions occur either spontaneously or upon interaction with a ligand. The MWC model postulates a thermal equilibrium between two distinct states while the KNF model postulates that conformational changes are induced by binding events. These models have proved successful at fitting experimental data and multiple extensions have been developed~\cite{szabo1972mathematical,herzfeld1974general,cui2008allostery}.
They leave, however, a fundamental question unanswered: what physical and evolutionary factors justify the presence of either one or several internal states?

Here, we address this question by studying a simplified physical model of protein allostery, where the presence of one or several states can emerge naturally due to physical and evolutionary constraints. Simplified physical models of proteins have been extensively studied in the context of protein folding, a problem that they helped to solve~\cite{onuchic2004theory}. More recently, several such models have been developed to study the physics and evolution of allostery~\cite{hemery2015evolution,yan2017architecture,tlusty2017physical,rocks2017designing,flechsig2017design,yan2018principles,dutta2018green,ravasio2019mechanics,rivoire2019parsimonious, rocks2019limits}. These models, however, constrain the mechanism to involve either one or two states. Here we introduce a more general model, which allows us to analyze the role that one or multiple states play in allostery.

The central result is that two archetypal mechanisms of cooperative allostery can arise, depending on the physical and selective constraints under which evolution takes place. First, a single-state mechanism  where ligand binding actuates a soft normal mode. Second, a multi-state mechanism where ligand binding stabilizes an alternate stable state, resulting in a switch-like conformational change. The former necessarily emerges when the energy landscape is constrained to be smooth. In contrast, when the energy landscape has the possibility to be rugged, we find that the second mechanism provides a statistically more likely and effective cooperative mechanism. We elucidate the origin of this evolutionary outcome and thus provide a testable explanation for the pervasiveness of multiple states in allosteric proteins.

\section{Model}

\begin{figure}
\centering
\includegraphics[width=.95\linewidth]{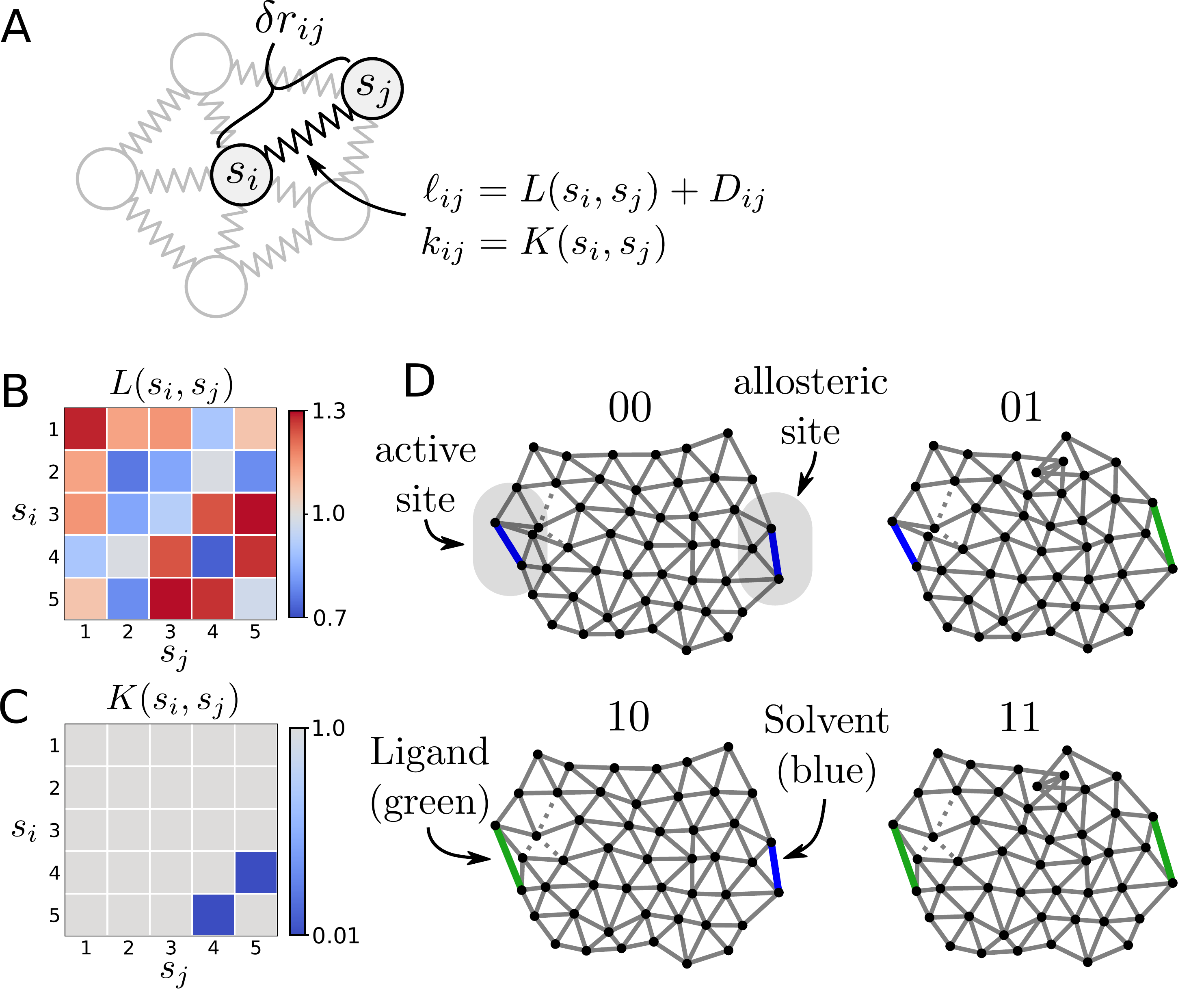}
\caption{The elastic network model -- (A) The physical parameters of the networks are determined by a sequence $s=(s_1,\dots,s_N)$ that specifies the type $s_i \in \{1,...,Q\}$ of each bead $i$. Beads are organized in a two-dimensional triangular lattice. The spring connecting beads $i$ and $j$ has stiffness $K(s_i, s_j)$ and rest length $L(s_i, s_j) + D_{ij}$ (see SI). (B) The table $L(s_i, s_j)$ has entries drawn uniformly in $[1-\s,1+\s]$ where $0<\s<1$ controls the disorder of the interactions. (C) The table $K(s_i, s_j)$ has all entries with $K(s_i,s_j)=1$ except for one soft interaction with $K(s_i,s_j)=0.01$. (D) Ground-state structures of an elastic network for the four possible ligand combinations. Solid lines represent stiff springs ($K(s_i, s_j)=1$) and dashed lines represent soft springs ($K(s_i, s_j)=0.01$). Ligand binding is modeled by changing the rest length of springs between two pairs of beads that define the active or allosteric sites. One value of the rest length defines a ligand (in green) and the other the solvent (in blue).}
\label{theModel}
\end{figure}

Our model abstracts proteins into a two-dimensional elastic network of beads whose interactions depend on their types, which can take $Q$ values playing the role of the 20 amino acids constituting protein sequences (Figure \ref{theModel}A).
The energy of a network with sequence $s=(s_1,\dots,s_N)$ and conformation $\bm{r}=(\bm{r}_1,\dots,\bm{r}_N)$ is of the form
\begin{equation} \label{eq:ham}
U(\bm{r},s) = \frac{1}{2}\sum_{ \langle i,j\rangle} k_{ij} \left(\delta r_{ij}- \l_{ij}\right)^2
\end{equation}
where the sum is over adjacent beads in the network and where $\delta r_{ij}=|\bm{r}_i-\bm{r}_j|$ is the distance between beads $i$ and $j$, $k_{ij}$ is the stiffness of the spring that connects them and $\ell_{ij}$ is its rest length. Both the values of $k_{ij}$ and $\ell_{ij}$ depend on the types $s_i$ and $s_j$:
\begin{equation} 
k_{ij} = K(s_i,s_j), \quad \l_{ij} = L(s_i,s_j)  + D_{ij}.
\end{equation}

To prevent unphysical behaviors, small amount of spatial disorder $D_{ij}$ is added to the rest lengths and an exclusion term is introduced between non-adjacent beads (see SI). The values of $K(s_i, s_j)$ and $L(s_i, s_j)$ are given by $Q\times Q$ interaction tables (Figure \ref{theModel}B-C). Specifically, we consider all entries in $K(s_i,s_j)$ to be stiff ($K=1$) except for one interaction that is soft ($K=0.01$). The entries of $L(s_i,s_j)$ are drawn uniformly in $[1-\s,1+\s]$ where $0<\s<1$ is a parameter controlling the disorder in the rest lengths. When $\s=0$, the absolute minimum of $U(\bm{r},s)$, which defines the ground state conformation, is substantially lower than other local minima associated with excited states. Increasing $\sigma$ adds frustration (the extent to which springs are stressed in the ground state) which decreases the energy gap between the ground state and excited states, thus resulting in a more rugged energy landscape (Figure \ref{landscapes}).

We define two springs on the opposite sides of the network to represent  active and allosteric sites. Binding at these sites is modeled by changing the rest length of the spring from one value representing the ``solvent'' $\l_0$ to another value representing the ``ligand'' $\l_1$ (colored bonds in Figure \ref{theModel}D; their stiffness is left unchanged, $K=1$). Four combinations of ligand binding are therefore possible, depending on whether the two sites are unbound (with ground-state energy denoted $E_{00}$), one site is bound (energies $E_{01}$ and $E_{10}$) or both sites are bound (energy $E_{11}$). These energies depend on the sequence $s$ and are estimated using a variant of the Basin-Hopping algorithm \cite{wales1997global} (Methods). We quantify allostery by the extent to which the binding energy at the active site depends on the presence of a ligand at the allosteric site, that is,
\begin{equation} \label{allo}
    \Delta \Delta E(s) = (E_{10}(s) - E_{00}(s) ) - (E_{11}(s) -  E_{01}(s)).
\end{equation}
We evolve the sequence $s$ of a network using a Monte Carlo algorithm with the analog of mutations consisting of randomly changing the type of the bead and a fitness defined by $\Delta\Delta E$ (Methods).

\begin{figure*}
\centering
\includegraphics[width=.95\linewidth]{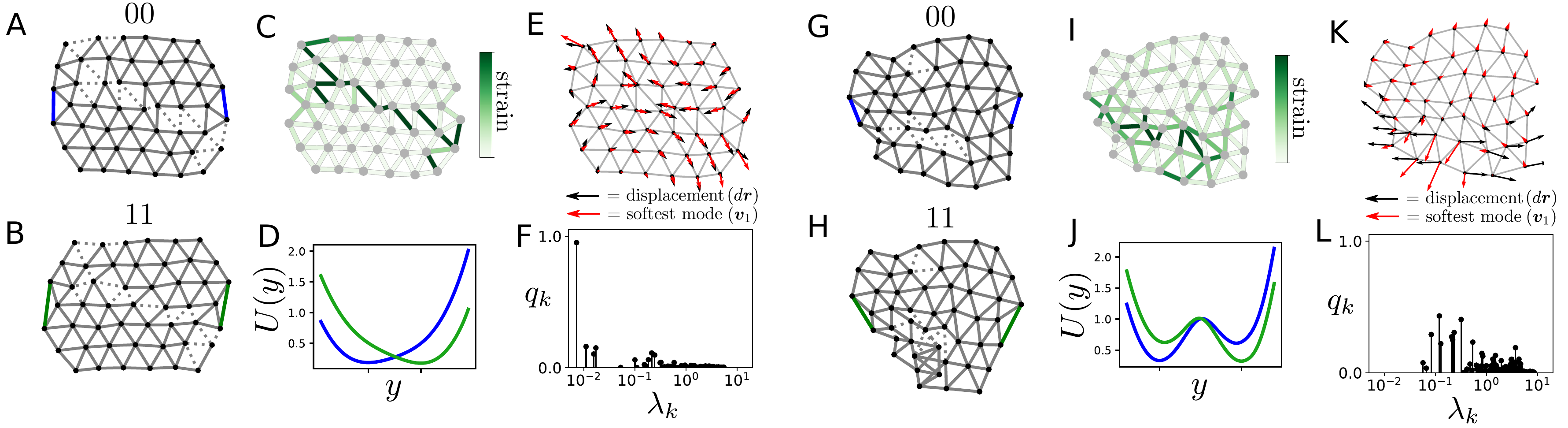}
\caption{Archetypes of networks with two different allosteric mechanisms -- (A-F) show a network evolved with homogeneous interactions ($\s=0$), which displays a single-state mechanism. (G-L) show a network evolved with disordered interactions ($\sigma=0.3$), which displays a two-state mechanism. (A-B) Ground state structures of the fully-solvated (00) and fully-bound (11) networks. (C) Bond strain between the structures in A and B. (D) Energy of the fully solvated (blue) and fully bound (green) networks along the conformational coordinate $y$ interpolating between the structures in A and B (Methods). (E) Structural displacements between the structures in A and B (00 $\rightarrow$ 11, in black) and motion of the softest mode of the unbound state (00, in red), showing that the two are aligned. (F) Overlap $q_k$ between the structural displacement upon binding ligand $d\bm{r}$ and each normal mode of the network $\bm{v}_k$ ($q_k = |\bm{v}_k \cdot d\bm{r}/\|d\bm{r}\| \ |$) as a function of the mode stiffness $\lambda_k$, which shows, as in E, a strong overlap along the first mode. (G-L) The same analyses for a network with a two-state mechanism, showing in contrast a major conformational change (H versus G), an energy with two minima along the conformational coordinate, and no overlap between the displacement induced by the ligand and the softest normal modes (K-L).}
\label{theMechs}
\end{figure*}

\section{Results}

\subsection*{Physical constraints on allostery}

Previous models of allostery based on elastic networks~\cite{yan2017architecture,rocks2017designing, dutta2018green, yan2018principles, flechsig2017design} had two key design properties. First, they had no or little frustration, the condition in which stressed springs occur in the ground state. Second, mutations changed the spring stiffness of interactions. Our model implements these design properties when $\sigma=0$.

In this limit, evolving networks for allostery reproduces the mechanism described in previous works~\cite{yan2018principles, tlusty2017physical, rocks2017designing} -- a soft normal mode connects the active site to the allosteric site and ligand binding induces a strain that aligns with the softest normal mode (Figure \ref{theMechs}A-F). This soft normal mode defines a direction along which the system can fluctuate substantially even without a large energetic input. The energy along the conformational coordinate upon ligand binding is indeed nearly flat with a single minimum (Figure \ref{theMechs}D), This mechanism defines ``single-state allostery''.

Making the energy landscape more rugged by taking $\sigma > 0$ qualitatively changes this picture (Figure \ref{theMechs}G-L). A network evolved for allostery in this context can now switch upon binding to a different stable state that is separated by an energy barrier (Figure \ref{theMechs}J). Additionally, the allosteric displacement no longer necessarily overlaps with the softest mode (Figure \ref{theMechs}L). This mechanism defines ``multi-state allostery''.

\begin{figure}
\centering
\includegraphics[width=.95\linewidth]{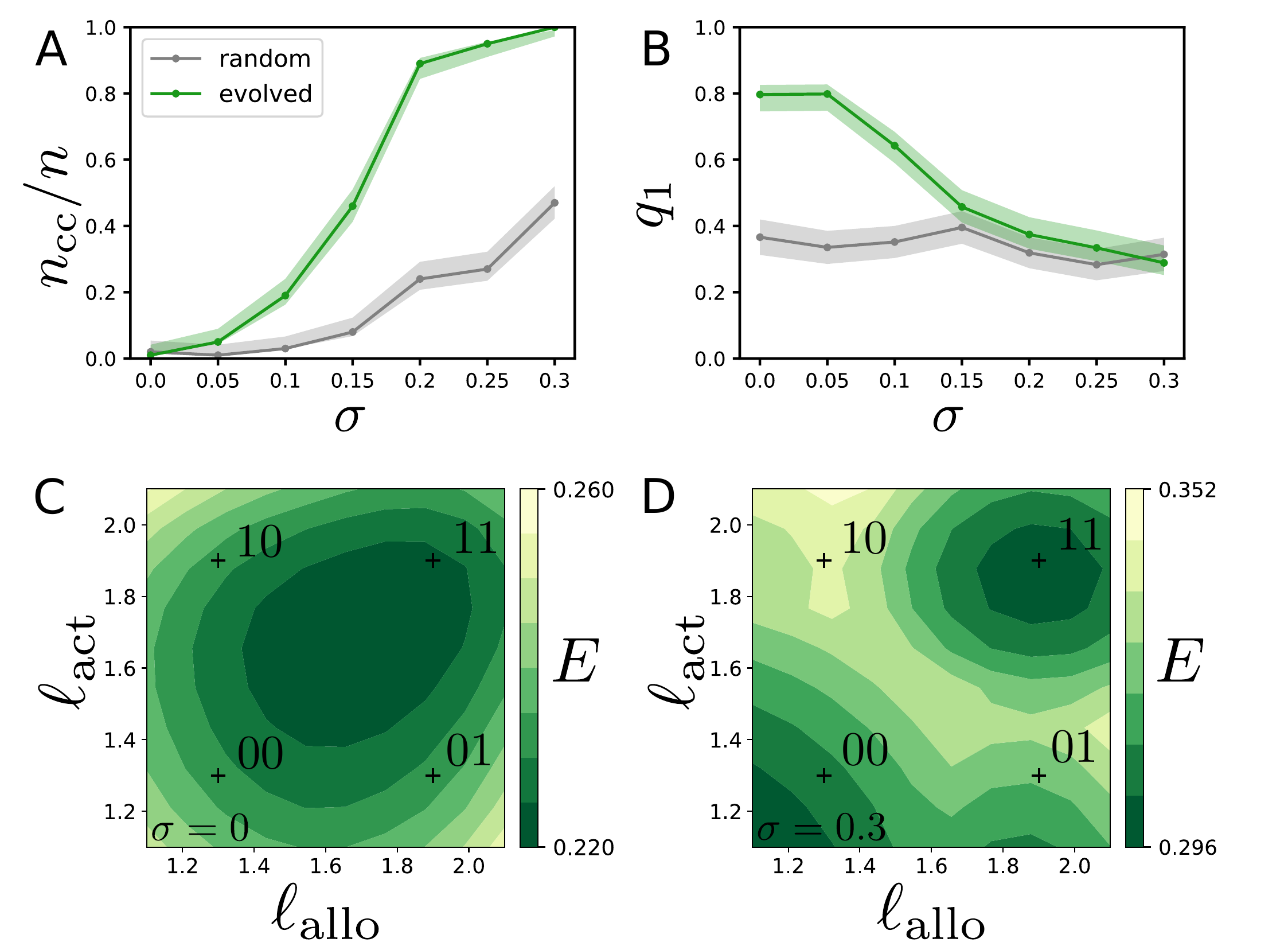}
\caption{Mechanism of allostery as a function of interaction disorder -- (A-B) Statistics over $n= 100$ networks evolved to maximize $\Delta\Delta E$ over 500 Monte Carlo iterations as a function of the disorder $\s$ of the interaction table $L(s_i,s_j)$. (A) Fraction of networks that undergo a multi-state conformational change $n_{cc}/n$ (errors bars are 95\% Wilson CI). (B) Mean overlap between the softest mode of the network $\bm{v}_1$ and the allosteric displacement upon ligand binding $d\bm{r}$ computed as $q_1 = |\bm{v}_1 \cdot d\bm{r}/\|d\bm{r}\| \ |$ (error bars are 95\% Wald CI). (C-D) Examples of ground state energy surfaces for networks evolved with homogeneous and disordered interactions respectively. The ``+'' marks the locations of different binding combinations.} 
\label{bigExp}
\end{figure}

Which of these mechanisms emerge over evolution  depends critically on the diversity of available interactions imposed by $\s$ : as $\s$ increases, the energy landscape becomes more rugged and the fraction of evolved networks with a multi-state conformational change ($n_{cc} / n$) increases while the overlap between the softest mode and the allosteric displacement decreases (Figure \ref{bigExp}A-B). This is also the case for random networks (gray curves in Figure \ref{bigExp}A-B) but evolved networks show a significant enrichment in multi-state mechanisms for large $\s$ and of single-state mechanisms for low $\s$ (green curves).  The results of Figure \ref{bigExp}AB are robust to parameter choice (Figure \ref{massive}).

The difference between mechanisms is also illustrated by representing the ground-state energy of evolved networks as a function of different binding ligands, corresponding to varying the rest lengths $\la$ and $\lb$ of the springs defining the active and allosteric sites (Figure \ref{bigExp}C-D). In this representation, the energy landscape associated with a single-state allosteric network typically takes the form of an anisotropic basin elongated in the $\l_{\rm act} = \l_{\rm allo}$ direction such that $E_{11}+E_{00}<E_{10}+E_{01}$. On the other hand, the energy landscape associated with a multi-state allosteric network can display two minima around $00$ and $11$, which also achieves $E_{11}+E_{00}<E_{10}+E_{01}$.

\subsection*{Evolutionary constraints on allostery}

\begin{figure}
\centering
\includegraphics[width=.95\linewidth]{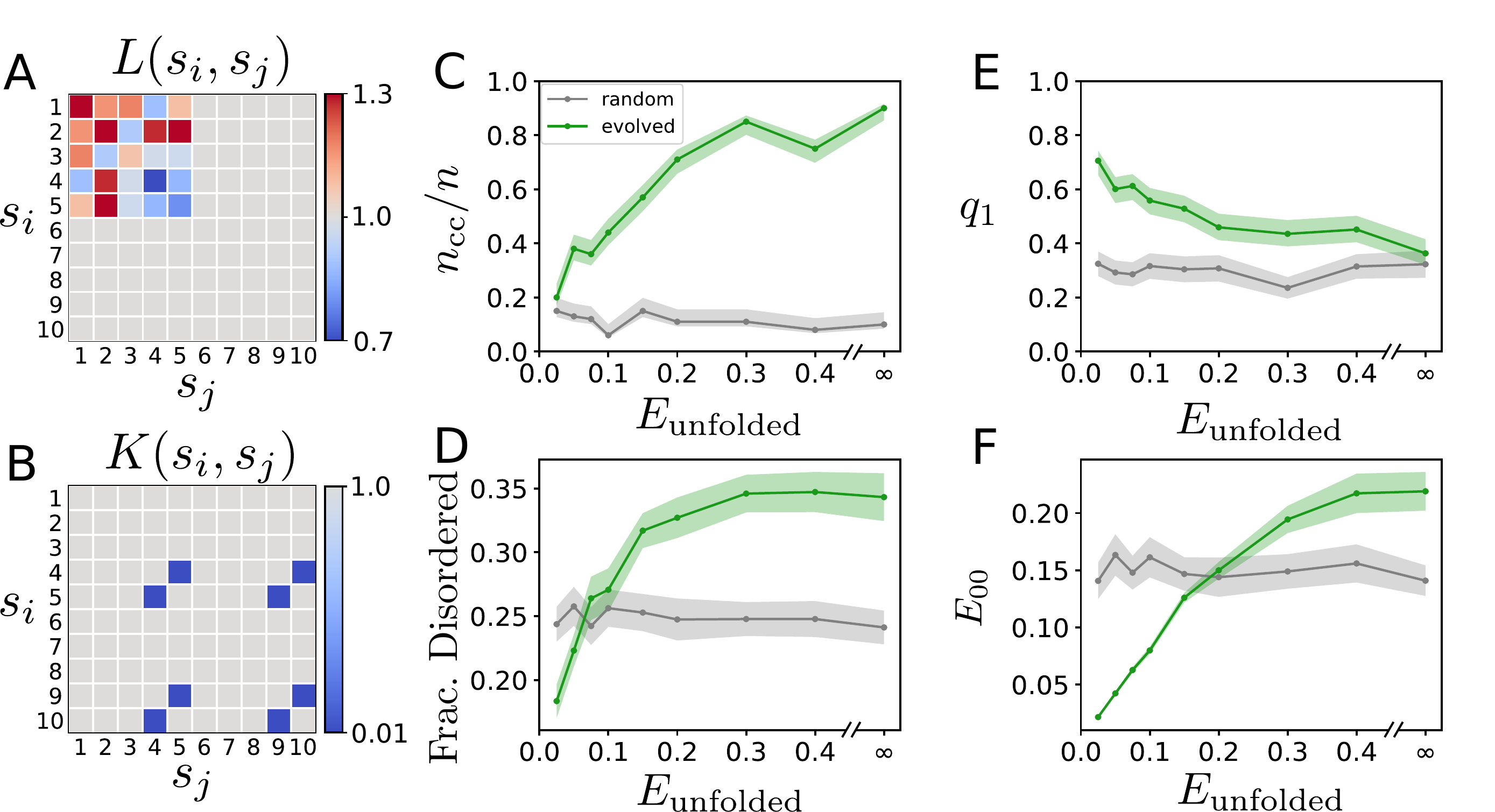}
\caption{Selection of allosteric mechanism -- (A) An example of a rest-length table $L(s_i,s_j)$ designed to contain both homogeneous and disordered interactions. (B)~Spring-constant interaction table $K(s_i,s_j)$.  (C-F) Properties of networks evolved under a joint pressure for cooperativity (high $\Delta\Delta E$) and stability (low $E_{00}$) as a function of the intensity $E_{\mathrm{unfolded}}$ of the selective pressure for stability (statistics over 100 networks evolved through 500 Monte Carlo iterations). (C) Fraction of networks that undergo a two-state conformational change (errors bars are 95\% Wilson CI). (D) Fraction of interactions between beads of type 1-5 ($(s_i,s_j) \in \{ 1,2,3,4,5 \}^2$; errors bars are 95\% Wilson CI). (E) Mean overlap of the allosteric displacement upon ligand binding (00$\rightarrow$11) and the softest non-zero mode of the network in the unbound state (errors bars are 95\% Wald CI). (F) Ground-state energy of the unbound state (errors bars are 95\% Wald CI). In absence of constraint for stability ($E_{\mathrm{unfolded}}=\infty$), networks evolve a multi-state mechanism using the disordered interactions. As the constraint for stability is increased ($E_{\mathrm{unfolded}}$ decreased), they tend to evolve a single-state mechanism relying on non-disordered interactions.}
\label{stability}
\end{figure}

The previous results show that the diversity of available physical interactions determines the mechanism of allostery, but if both mechanisms can evolve, which does evolution favor? We address this question by considering a set of interactions that contains both homogeneous ($\sigma=0$) and disordered ($\sigma=0.3$) rest length distributions (Figure \ref{stability}A). Networks evolved with such an interaction table have effectively the ``choice'' to populate a part of sequence space with either smooth or rugged local energy landscapes. Simulations show that evolution under selection for allostery leads to networks with a greater number of disordered interactions  (those  between beads of type 1-5) than those of random networks (random: 24\%, evolved: 34\%). Correspondingly, the vast majority of these evolved networks are poised to display multi-state conformational changes (random: 10\%, evolved: 90\%) and an overlap with the softest mode comparable to those of random networks (random: $q_1= 0.32$, evolved: $q_1= 0.36$). These results are represented by the data-point $E_{\rm unfolded}=\infty$ in Figure~\ref{stability}. They indicate that the multi-state mechanism is more competitive than the single-state mechanism. That is, when the interactions provide sufficient disorder, networks tend to evolve towards multi-state allostery.  

In real proteins, the analog of our interaction tables are interactions between the amino acids. The physics of these interactions are largely fixed and $\s$ is therefore not subject to physiological control. But, some tunable parameters may play a role similar to $\s$ and effectively control the ruggedness of the energy landscape. One such parameter is thermal stability, which is itself subject to natural selection. As a proxy for thermal stability, we introduce here an arbitrarily defined energy $E_{\rm unfolded}$ to represent the energy of a non-functional unfolded state. To impose a stability constraint, we again select networks based on $\Delta \Delta E$ but now restrict the evolution to sequences satisfying $E_{00} < E_{\rm unfolded}$.

By evolving networks for varying values of $E_{\mathrm{unfolded}}$, we verify that the smaller $E_{\mathrm{unfolded}}$ is, the more stable the networks are (Figure \ref{stability}F). The most stable networks are less likely to have multi-state conformational changes (Figure \ref{stability}C) and, on average, show a greater overlap between the allosteric displacement and the softest mode (Figure \ref{stability}E). Additionally, the usage of disordered interactions decreases with decreasing $ E_{\mathrm{unfolded}}$ (Figure \ref{stability}D). These results show how an additional selective pressure, here stability, can alter the likelihood to evolve a single or multi-state mechanism of allostery. The underlying phenomenon is the same as before -- stable networks minimize frustration, and thus tend to populate smoother parts of the landscape, effectively corresponding to a smaller $\s$. Thus, thermal stability is a parameter that can, in principle, control the mechanism of allostery.

\subsection*{1-D model}

\begin{figure}
\centering
\includegraphics[width=.9\linewidth]{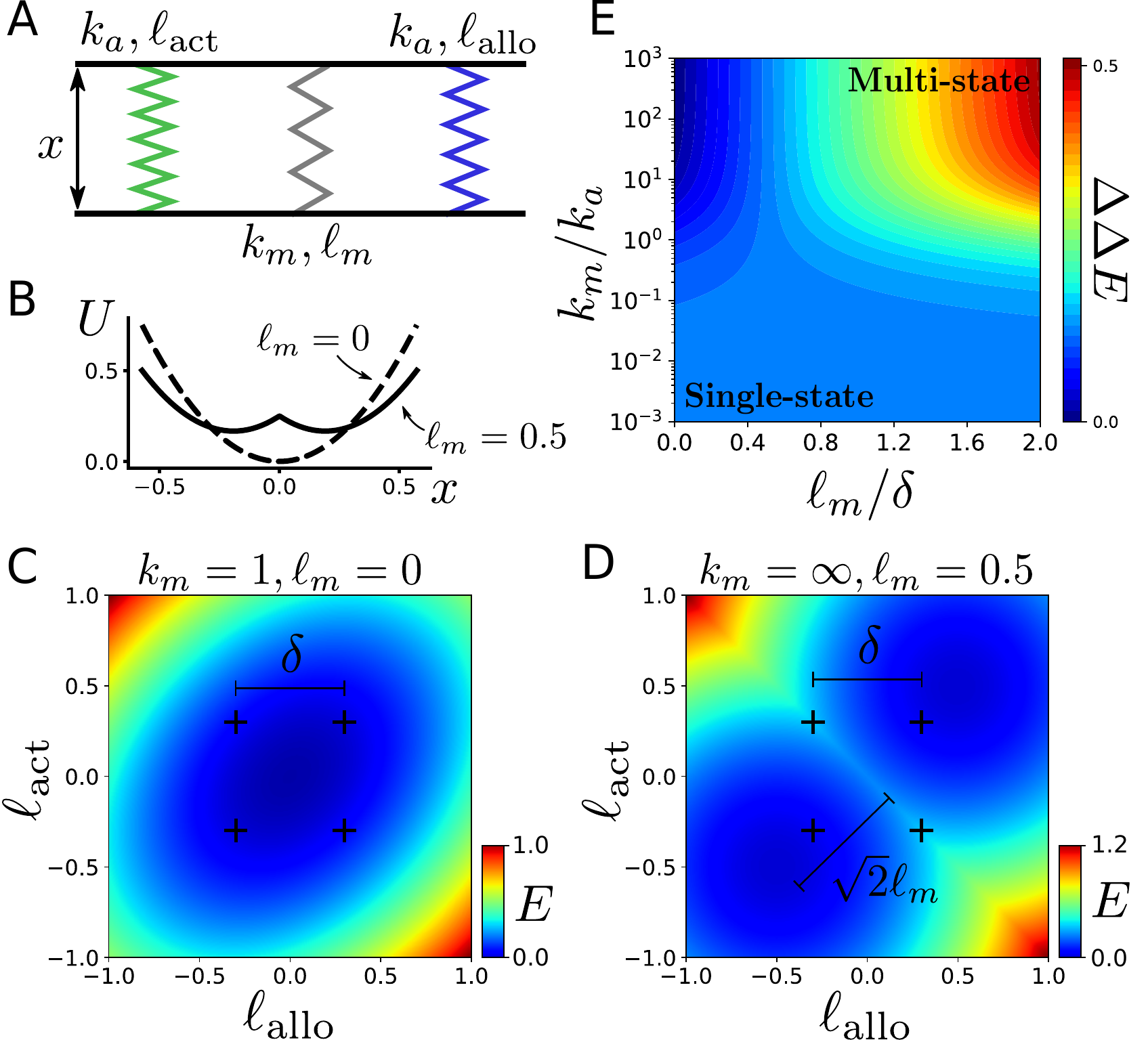}
\caption{1d model -- (A) The model consists of three springs connecting two rigid bars constrained to move in one dimension. Two springs are analogous to the ligand springs of the elastic network. The third spring, whose contribution to the energy is $U_m(x)=k_m(|x|-\l_m)^2/2$, mediates the mechanism of allostery and can flip. (B) Total energy $U(x)$ as function of $x$ for two values of $\l_m$: the energy landscape has a single minimum $\ell_m=0$ and two minima when $\ell_m>0$. (C) Small values of $k_m,\l_m$ correspond to a single-state mechanism. (D) Large values of $k_m,\l_m$ correspond to a multi-state mechanism. (E) Cooperativity $\Delta\Delta E$ as a function of the normalized quantities $k_m/k_a$ and $\l_m/\delta$, showing a continuum between purely single-state and strongly two-state mechanisms. }
\label{curve}
\end{figure}

The two mechanisms displayed by our two-dimensional network model can be illustrated in an even simpler one-dimensional model. This model consists of two rigid bars connected by three springs -- two harmonic springs represent the active and allosteric sites and a single elastic spring mediates the mechanism of allostery (Figure \ref{curve}A). The energy of this model is
\begin{equation}
    U(x) =  \frac{1}{2}\Big[k_a(x-\ell_{\mathrm{act}})^2
        + k_{m}(|x|-\ell_{m})^2
    + k_a (x-\ell_{\mathrm{allo}})^2 \Big].
\end{equation}

The ground-state energy $E(\ell_{\mathrm{act}} , \ell_{\mathrm{allo}})=\min_x U(x)$ is harmonic with a single minimum when $\l_m=0$ and has two minima when $\ell_m>0$ (Figure \ref{curve}CD). For simplicity, we assume that the rest lengths at the two binding sites take the same two (algebraic) values $\l_0$ and $\l_1$, with $\l_0 = - \l_1$. Under these assumptions, the cooperativity is (see Methods)
\beq \label{1d_cooperativity}
\Delta\Delta E=\frac{k_a\left(k_a\d+2k_m\l_m\right)\d}{2k_a+k_m},
\eeq
with $\d = | \l_1 - \l_0 |$.
The mechanism is strictly single-state only for $\l_m=0$, in which case $\Delta\Delta E$ cannot exceed $k_a\d^2/2$. On the other hand, when $\l_m>0$, $\Delta\Delta E$ can take arbitrarily large values and scales as $\Delta\Delta E\sim k_a\l_m\d$ when $k_m\to\infty$. A continuum of mechanisms exists between these two extremes (Figure \ref{curve}E) but, when viewing $\l_m$ as a physical property of the landscape and $k_m$ as an evolutionary parameter, two distinct regimes emerge: when $\l_m<\delta/4$, a single-state mechanism with a soft mode ($k_m=0$) is optimal while when $\l_m>\delta/4$, a two-state mechanism with a large barrier ($k_m=\infty$) is optimal. Finally, if $\l_m$ is itself subject to evolution, maximal cooperativity $\Delta\Delta E$ is achieved by the two-state mechanism ($\ell_m=\infty$, $k_m=\infty$).

By defining analogs of $k_m$ and $\ell_m$, we verify that the two-dimensional model behaves indeed as the one-dimensional model (Figure \ref{validation1dModel}).

\section{Discussion}

In this work, we study the evolutionary origins of single or multiple states in systems selected for allostery using a minimal yet generic physical model of protein allostery. Depending on the ruggedness of the underlying energy landscapes, two archetypal mechanisms emerge: in smooth landscapes, a single-state mechanism evolves while in rugged landscapes a multi-state mechanism evolves. In single-state systems, allostery is mediated through actuation of a soft global mode induced by ligand binding whereas in multi-state systems ligand binding stabilizes an alternate pre-existing state. We find that the multi-state mechanism has a greater potential for cooperativity and is therefore evolutionarily favored whenever it has the possibility to evolve. We also find that additional selective pressures can modulate the outcome of evolution; in particular, a strong selection for stability favors the evolution of single-state mechanisms. These results are demonstrated by numerical simulations in a two-dimensional elastic network model and recapitulated by analytical calculations in a simpler one-dimensional model.

For computational tractability, we limited our simulations to zero temperature and ignored entropic effects, on which allostery can also rely~\cite{cooper1984allostery}. However, the distinction between single-state and multi-state mechanisms extends to finite temperature by considering the free energy instead of the ground-state energy. We carried out this generalization for the one-dimensional model, showing that temperature does not modify our general conclusions (SI text, Figure \ref{temperature}). Our model also does not explicitly account for partial or global unfolding ~\cite{mitrea2013regulated} but we note that allosteric mechanisms where ligand binding differentially stabilizes the folded and unfolded ensembles are instances of multi-state mechanisms provided that folding is a cooperative process \cite{hilser2007intrinsic}.  

The distinction between single and multi-state mechanisms of allostery reflects the two classes of phenomenological models that have been introduced to describe allostery, namely the MWC and KNF models~\cite{monod1965nature,koshland1966comparison}. The KNF model, where state changes only upon ligand binding, assumes a single-state mechanism, while the MWC model, where two states preexist ligand binding, assumes a two-state mechanism. Indeed, the physical model described here can be reformulated from a thermodynamic perspective to be analogous to MWC and KNF models (SI text). The MWC and KNF models are archetypes at the two ends of a continuum of models for allostery that interpolate between them~\cite{herzfeld1974general}. Thus the work presented here provides a foundation for understanding the assumptions underlying these models, and perhaps more importantly, the evolutionary origin of specific models for allostery. 

Our results also stand in contrast to previous works with simplified physical models of allostery. Many of these previous works assumed as we do an elastic network but restricted their analyses to linear responses and therefore did not allow for multi-state allostery ~\cite{yan2017architecture, rocks2017designing, dutta2018green, ravasio2019mechanics}. Elastic network models have also been extensively studied as coarse-grained models for studying thermal fluctuations based on crystal structures of proteins \cite{bahar1997direct,atilgan2001anisotropy}. In our work, however, elastic networks are only used to implement a generic continuous energy landscape and we expect our results to extend to other classes of potentials, provided their ruggedness can be tuned.

A critical parameter of our model is the diversity $\s$ of available interactions -- the central factor controlling the ruggedness of the fitness landscape. This raises the question of what value of $\s$ is representative of the interactions between amino acids in actual proteins. The fact that conformational switching between stable states has been observed in a large range of proteins~\cite{frauenfelder1988conformational,volkman2001two,james2003antibody} suggests that amino acid interactions are indeed diverse enough to permit the evolution of multi-state allostery. The finding that most proteins are only marginally stable~\cite{bloom2004stability} also suggests a relatively weak selective pressure for thermal stability, leaving room for multiple states. Significant frustration, which $\s>0$ induces in our model, has in fact been estimated in proteins ~\cite{ferreiro2007localizing, ferreiro2011role}. Taken together, the data suggest that proteins selected to maximize allosteric cooperativity are statistically more likely to work through a multi-state mechanism.

Indeed, this statement is consistent with the diversity of mechanisms underlying allostery in proteins. Previous analyses estimating the largest overlap between the allosteric displacement and any normal mode ($\mathrm{max}_i(q_i)$) of allosteric proteins have found a broad range of values ~\cite{ravasio2019mechanics, tama2001conformational, zheng2006low}. As predicted, most are far from archetypal single-state allostery where $q_1 \simeq 1$, but the quantitative diversity of mechanisms remains to be explained. It could arise from additional selective pressures that tune the evolved mechanism of allostery, or from kinetic constraints not accounted for by our model. For instance, energetic barriers for conformational switching cannot be too large or the time to cross them becomes prohibitive~\cite{ravasio2019mechanics}.

In this work, we focus on the emergence of multiple states in proteins with direct selection for allosteric regulation. However, allosteric effects in proteins have been proposed to precede their use for regulation \cite{lee2008surface, coyle2013exploitation, reynolds2011hot}, raising the question of what is the origin of allostery in proteins. Rather than a consequence of direct allosteric selection, the presence of a conformational switch with allosteric effects may first arise as a by-product of selective pressures unrelated to regulation or signal transmission per se, and then later co\"opted for these purposes. Two non-exclusive scenarios are a selection for binding specificity~\cite{rivoire2019parsimonious} and a selection for evolvability~\cite{raman2016origins, murugan2021roadmap}. Our model provides a flexible platform to test these hypotheses and advance towards a general theory for the origin of models for allostery in proteins.


\begin{acknowledgments}
We thank Kabir Husain, Riccardo Ravasio, and members of the Ranganathan and Rivoire Labs for helpful discussions. RR and OR are grateful for funding from the FAACTS program of the France Chicago Center. RR acknowledges support from NIH RO1GM12345 and RO1GM141697, and the University of Chicago, and OR from ANR-17-CE30-0021-02.
\end{acknowledgments}

\bibliography{multistates}

\clearpage
\onecolumngrid

\appendix
\section{Methods}

\begin{figure}
\centering
\includegraphics[width=.8\textwidth]{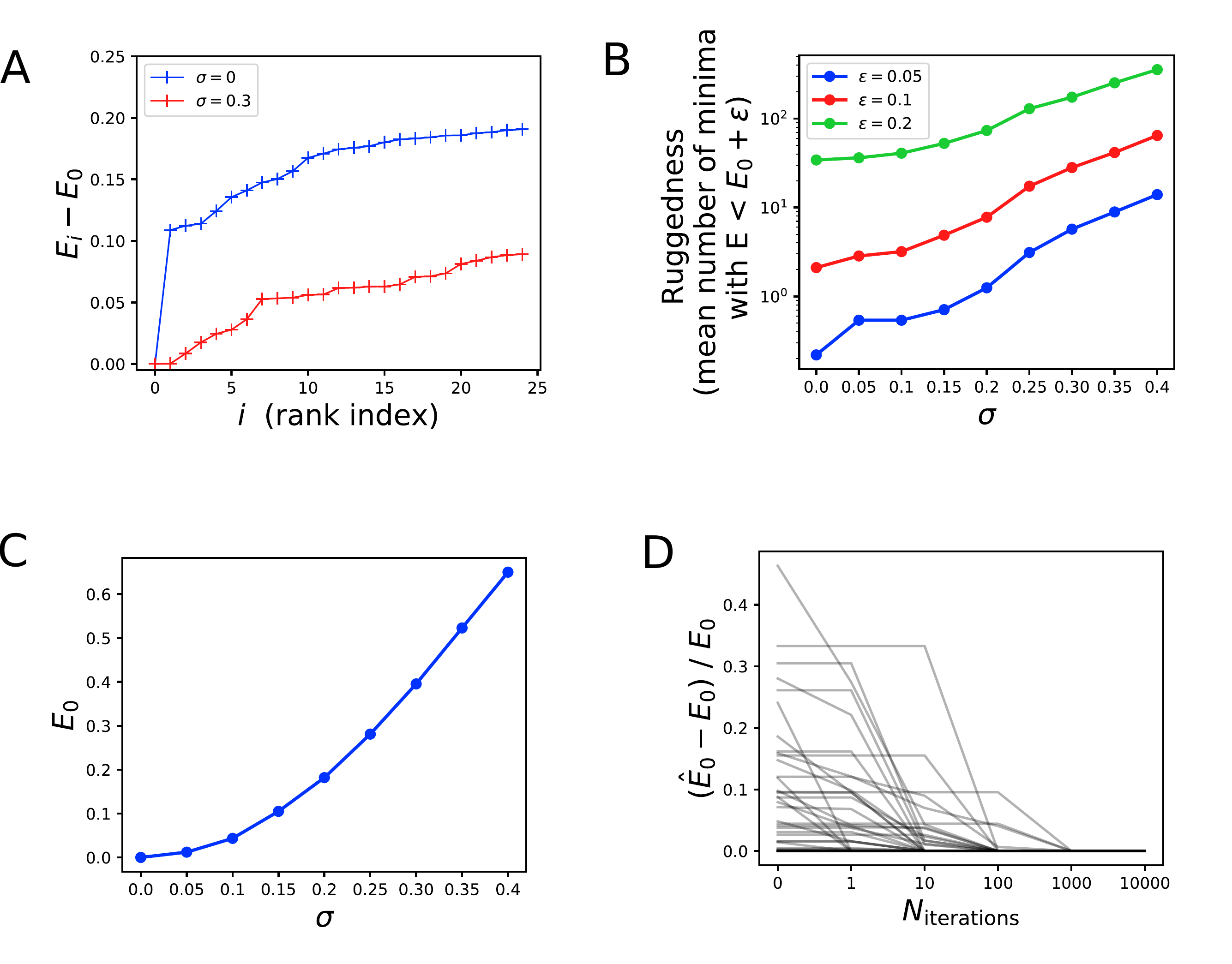}
\caption{Ruggedness and frustration increase with $\sigma$. (A) The energies of the 25 lowest minima relative to the ground state energy ($E_0$) for a network with $\s = 0$ (blue) and $\s = 0.3$ (red). The networks with ordered interactions ($\s = 0$) have a ground state far more stable than any other minima, whereas networks with disordered interactions $\sigma = 0.3$ have many near-degenerate low energy states. (B) We measure ruggedness of an energy landscape as the number of minima within $\epsilon$ of the ground state energy ($E_0$). For each $\s$, we plot the ruggedness averaged over 100 random networks. Higher $\s$ implies more low energy minima relative to the ground state. Minima are found with the ground-state finding algorithm (Methods). (C) Frustration, the extent to which bonds are stressed, is measured by the ground state energy $E_0$. A ground state where all springs are relaxed has $E_0=0$. For each $\s$, 100 random networks are generated and their ground states are estimated with the ground-state finding algorithm (Methods). (D) The error in the estimated ground state energy $\hat{E}_0$ for increasing number of iterations of the ground state finding algorithm for 100 random networks. ($\sigma = 0.3$ and population size $p=20$). At 100 iterations, 96\% of ground states estimates are correct. The energies after 10000 iterations are taken to be the true ground state energies, $E_0$. These data show that the ground state can be estimated with sufficient accuracy even when the energy landscapes are rugged with many minima.}
\label{landscapes}
\end{figure}

\begin{figure}
\centering
\includegraphics[width=\textwidth]{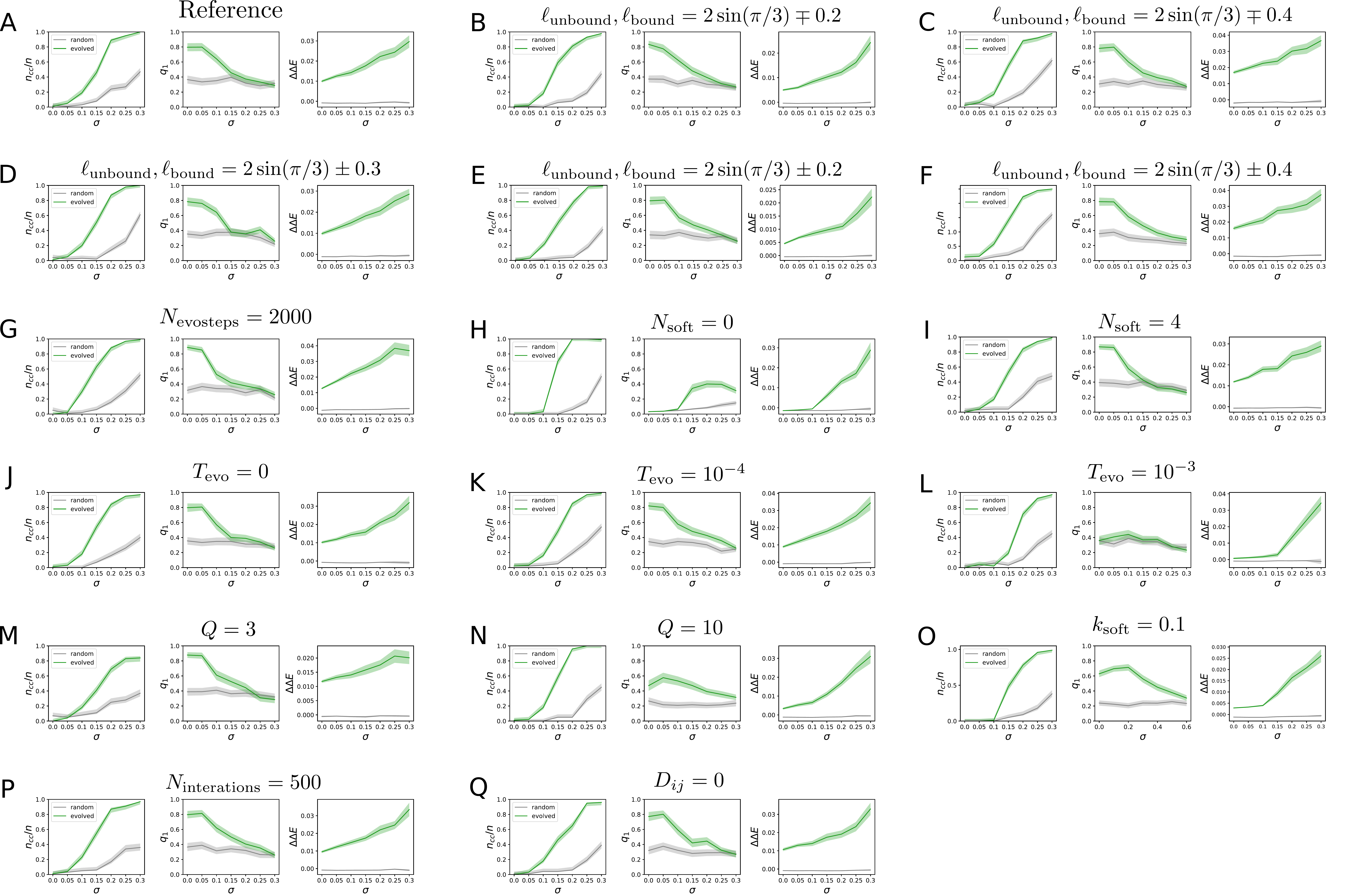}
\caption{Robustness of Fig 3AB with respect to various parameters. Each panel shows the fraction of networks that undergo a multi-state conformational change upon binding ligand $n_{\mathrm{cc}} / n$, the mean overlap of the allosteric mode with the softest mode $q_1$, and the mean cooperativity $\Delta \Delta E$ for 100 random (gray) and evolved (green) networks. In panels B-Q only one parameter is changed from the reference in panel A. (A) Same data as  in Figure 3AB where the parameters of the simulation are: $\l_0=2\sin (\pi/3)-0.3$, $\l_1=2\sin(\pi/3)+0.3$,
$N_{\mathrm{evosteps}} = 500$, $N_{\mathrm{soft}} = 2$, $T_{\mathrm{evo}} = 10^{-5}$, $Q=5$, $k_{\mathrm{soft}} = 0.01$,$N_{\mathrm{iterations}} = 100$, and $D_{ij}$ as defined in Methods. (B-F) Results do not significantly change with different choices of ligands except that cooperativity decreases when the difference between $\l_0$ and $\l_1$ decreases, as predicted by the 1d model. (G) Results do not change when the number of interactions of the Monte Carlo evolution is increased from $N_{\mathrm{evosteps}} = 500$ to $N_{\mathrm{evosteps}} = 2000$. (H) When there are no soft interactions permitted between beads, $N_{\mathrm{soft}}=0$, (i.e. if $K(s_i, s_j)=1$ for all entries) no cooperativity evolves at low $\s$. (I) Doubling the number of soft interactions from  $N_{\mathrm{soft}}=2$ to $N_{\mathrm{soft}}=4$ does not significantly change the results. (JKL) Changing the temperature of the evolutionary Monte Carlo $T_{\mathrm{evo}}$ does not change the results as long as it is sufficiently small, $T_{\mathrm{evo}} <10^{-3}$. At $T_{\mathrm{evo}} \geq 10^{-3}$ nearly no cooperativity evolves at small $\s$. This result was previously described in\cite{yan2017architecture}. (MN) The results are unchanged with different number of bead types $Q$. (O) The results are unchanged when the softness of the soft interactions is increased to $k_{\mathrm{soft}} =0.1$. (P) The results are unchanged if the number of iterations of the basin hopping algorithm is increased from $N_{\mathrm{iterations}} = 100$ to $N_{\mathrm{iterations}} = 500$. (Q) The results are unchanged when the spatial disorder is removed, $D_{ij}=0$.}
\label{massive}
\end{figure}

\noindent \textbf{Estimating ground states.} We estimate the ground state of a network with a version of the Basin-Hopping algorithm~\cite{wales1997global} where a genetic algorithm is used instead of a Monte Carlo algorithm to search conformation space. The algorithm (1) initializes a population of $p$ copies of a network; (2) perturbs ($\bm{r} \rightarrow \bm{r} + \bm \eta, \ \bm \eta \sim \mathcal{N}$) and relaxes each structure using the FIRE algorithm~\cite{bitzek2006structural}; (3) removes the $p/2$ structures with the highest relaxed energy; (4) replicates each remaining structure; (5) repeats steps 2-4 $N_{\rm{iterations}}$ times; (6) outputs the structure with lowest energy. We take $p=20$ and $N_{\mathrm{iterations}}=100$, which we find to be sufficient for estimating the ground state (Figure \ref{landscapes}D and \ref{massive}P).

\noindent \textbf{Definition of multi-state mechanisms.} Let $\bm{r}_{00}$ and $\bm{r}_{11}$ denote the ground-state conformations with no ligand (00) and two ligands (11). We define $\bm{r}_{00}'$ as the structure relaxed from $\bm{r}_{11}$ in absence of ligand and $\bm{r}_{11}'$ as the structure relaxed from $\bm{r}_{00}$ in presence of the two ligands. We say that the mechanism is single-state if $\bm{r}_{00} = \bm{r}_{00}'$ and $\bm{r}_{11} = \bm{r}_{11}'$, and multi-state otherwise.

\noindent \textbf{Normal modes.} Normal modes are computed by eigenvalue decomposition of the Hessian of the energy function in the unbound state. Three zero modes correspond to a global rotation $\bm{v}_{\mathrm{rot}}$ and two global translations $\bm{v}_{x}$ and $\bm{v}_{y}$. The nonzero modes and their stiffnesses are denoted $\bm{v}_k$ and $\lambda_k$ with $\lambda_i \leq \lambda_j$ if $i>j$, with $\bm{v}_1$ being the softest nonzero mode.

\noindent \textbf{Allosteric displacement.} Given $\bm{\Delta} = \bm{r}_{11} - \bm{r}_{00}$, the allosteric displacement $d \bm{r}$ is computed as
\beq 
d \bm{r} = \bm{\Delta} - (\bm{\Delta} \cdot \bm{v}_{\mathrm{rot}})\bm{v}_{\mathrm{rot}} - (\bm{\Delta} \cdot \bm{v}_{x})\bm{v}_{x} - (\bm{\Delta} \cdot \bm{v}_{y})\bm{v}_{y}.
\eeq

\noindent \textbf{Overlaps.} The overlap between a nonzero normal mode $\bm{v}_k$ and the allosteric displacement $d\bm{r}$ is computed as
\begin{equation}
    q_k = \left| \frac{d\bm{r}}{\|d\bm{r}\|} \cdot \bm{v}_k \right|.
\end{equation}

\noindent \textbf{Conformational coordinate.} The conformational coordinate $y$ is defined by $\bm{r}(y) = \bm{r}_{00} + y d\bm{r}$ and its associated energy by $U(y) = U(\bm{r}(y), s)$.

\noindent \textbf{Monte Carlo evolution.}  Networks are ``evolved'' using a Metropolis Monte Carlo method with an error catching step due to the error prone nature of the ground state finding algorithm. A candidate sequence $s'$, one mutation away from the current sequence $s$, is accepted with probability $p = \exp((  \Delta \Delta E (s') - \Delta \Delta E(s )) / T_{\rm evo})$ where $T_{\rm evo}=10^{-5}$. Following the acceptance or rejection step, the ground states of the current sequence, $E_{00}$, $E_{10}$, $E_{01}$, $E_{11}$ are systematically revaluated. If a lower ground state energy is found, the current sequence is set to the previous sequence and ground states are recomputed. This step prevents inaccurate ground state estimates from causing fictitiously large values of $\Delta \Delta E$.

\section{Supplemental Information}

\subsection{Elastic networks model details}

The full energy of a network with sequence $s=(s_1,\dots,s_N)$ and conformation $\bm{r} =(\bm{r}_1,\dots,\bm{r}_N)$ is
\begin{equation} \label{fullham}
U(\bm{r},s) = \frac{1}{2}\sum_{i>j} k_{ij} (\delta r_{ij}- \ell_{ij} )^2 + \frac{1}{2} k_{\rm rep} \sum_{i>j} (1-A_{ij}) \Theta(\ell_{\rm rep} -  \delta r_{ij}) (\delta r_{ij} - \ell_{\rm rep}) ^2,
\end{equation}
where $\delta r_{ij}=|\bm{r}_i-\bm{r}_j|$ is the distance between beads $i$ and $j$, $k_{ij}$ is the stiffness of the spring that connects them and $\ell_{ij}$ is its rest length. Both the values of $k_{ij}$ and $\ell_{ij}$ depend on the types $s_i$ and $s_j$:
\begin{equation} 
k_{ij} = A_{ij} K(s_i,s_j), \quad \l_{ij} = L(s_i,s_j) + D_{ij}.
\end{equation}
The values of $K(s_i, s_j)$ and $L(s_i, s_j)$ are given by $Q\times Q$ interaction tables. Specifically, we consider all interactions in $K(s_i,s_j)$ to be stiff ($K=1$) except for one randomly chosen entry that is soft ($K=0.01$). The entries of $L(s_i,s_j)$ are draw uniformly in $[1-\s,1+\s]$ where $0<\s<1$ is a parameter controlling the disorder in the rest lengths. The adjacency matrix $A_{ij}$ is designed by placing $N$ beads on a triangular lattice and taking $A_{ij}=1$ for nearest neighbors and $A_{ij}=0$ otherwise ($A_{ii}=0$). 

$D_{ij}$ adds spatial disorder to avoid possible complications from straight lines of beads \cite{yan2017architecture}.  

After beads are first placed on a triangular lattice to construct the adjacently matrix as described above, the pairwise distances are stored in $W^{\mathrm{lattice}}_{ij}$. The beads are then displaced by a random normal perturbation $\Delta \bm{r}_i \sim \mathcal{N}(\bm{\mu }, \mathbf{\Sigma})$ where ${\bm{\mu }}=\left[{\begin{smallmatrix}0\\0\end{smallmatrix}}\right]$ and $\mathbf{\Sigma}=\left[{\begin{smallmatrix}0.05&0\\0&0.05\end{smallmatrix}}\right]$. From this conformation, new pairwise distances are stored in $W^{\mathrm{pert}}_{ij}$ and $D_{ij} = W^{\mathrm{pert}}_{ij} - W^{\mathrm{lattice}}_{ij}$. The key feature of $D_{ij}$ is that it does not add any frustration to the network (i.e., when $\sigma=0$ the ground state energy $E_0=0$). Both $A_{ij}$ and $D_{ij}$ are fixed during evolution of a given network. 

To prevent unphysical networks collapse behaviors, the second term of Eq. [\ref{fullham}] defines a harmonic repulsion term between non-adjacent beads. $\Theta$ is the Heaviside function so that $ \Theta(\l_{\rm rep} -  \delta r_{ij})=1$ if $\l_{\rm rep} >  \delta r_{ij}$ and 0 otherwise. $\l_{\rm rep}=0.7$ and $k_{\rm rep} = 2$ are the distance cutoff and spring constant for the harmonic repulsion term, respectively. All networks simulated in this work have a size of $N=49$ beads.

\subsection{$\tilde k_m$ and $\tilde \l_m$}
In the 1d model, $k_m$ can be interpreted as the stiffness of the allosteric mode. To define an equivalent quantity for the 2d elastic network model, we consider $\Tilde{k}_m = \lambda_1 / \tilde{\lambda}$, where $\lambda_1$ is the stiffness of the softest non-zero mode and $\tilde{\lambda}$ is the median stiffness. In the 1d model, $\l_m$ is related to the height of the barrier of $U(x)$ when $k_a = 0$: $\Delta U_{\rm barrier} =  k_m \l_m^2 / 2 $. To define an equivalent quantity for
the 2d elastic network model, we consider $\Tilde{\l}_m = \sqrt{ 2 \Delta U_{\rm barrier} / \Tilde{k}_m}$ when $k_{\rm act} = k_{\rm allo} = 0$. $\Delta U_{\rm barrier} = 0$ if $U(y)$ has one minimum ($U'(y)=0, U''(y)>0$) and $\Delta U_{\rm barrier} = U(y_3) - {\rm max} (U(y_1),U(y_2))$, if $U(y)$ has two minima at $y_1$ and $y_2$ and one maximum at $y_3$. Other possibilities are not observed.

\subsection{1d model at zero temperature}

For a system of harmonic springs, the energy takes the form,
\beq\label{eq:syst}
V(x)=\frac{1}{2}\sum_ik_i(x-\l_i)^2=\frac{1}{2}\left(\sum_ik_i\right)\left(x-\frac{\sum_ik_i\l_i}{\sum_ik_i}\right)^2+V_{\rm min}
\eeq
where
\beq
V_{\rm min}=\min_x V(x)=\frac{1}{2}\left(\sum_ik_i\l_i^2-\frac{(\sum_ik_i\l_i)^2}{\sum_ik_i}\right).
\eeq
The 1d model of allostery is given by
\beq \label{eq:model}
U(x)=(1/2)k_a(x-\la)^2+(1/2)k_a(x-\lb)^2+(1/2)k_m(|x|-\l_m)^2.
\eeq
Putting \eqref{eq:model} in the form of \eqref{eq:syst} gives
\beq
V(x,\l_m)  = \frac{1}{2}(2k_a+k_m)\left(x-\frac{k_a(\la + \lb) + k_m \l_m}{2k_a+k_m}\right)^2 + V_{\rm{min}}(\l_m)
\eeq
where
\beq
V_{\rm{min}}(\l_m)  = \frac{1}{2} \left[ k_a(\la^2 + \lb^2) +k_m\l_m^2 - \frac{(k_a(\la + \lb) +k_m\l_m)^2}{2k_a+k_m} \right].
\eeq
The ground state of $U$ is defined by $E = \min_xU(x)=\min_{\pm }V_{\rm min}(\pm\l_m)$. This gives
\beq
E(\la,\lb) = \frac{k_a}{2(2k_a+k_m)}\left(k_a(\la-\lb)^2+k_m(\la^2+\lb^2+2\l_m^2)-2k_m|(\la+\lb)\l_m|\right)
\eeq
Cooperative allostery is computed as
\beq
\Delta\Delta E=E(\l_1,\l_0)+E(\l_0,\l_2)-E(\l_1,\l_2)-E(\l_0,\l_0).
\eeq
If $\l_1 = \l_2$ then, 
\beq
\Delta\Delta E = \frac{k_a}{2 k_a +k_m}\left( k_a|\l_1 - \l_0|^2 + 2k_m |\l_m| (|\l_1| +     |\l_0| - |\l_1+\l_0|) \right)
\eeq
If $\l_1 = -\l_0$ and $\d = |\l_1 - \l_0|$,
\beq \label{1dtheory}
\Delta\Delta E = \frac{k_a ( k_a\d + 2k_m \l_m)\d}{2 k_a +k_m}.
\eeq
The mechanism changes from single-state to multi-state at the transition rest length $\l_m^* = \d/4$. When $\l_m < \d/4$, $\Delta\Delta E$ is optimized at $\hat{k}_m=0$ and can not exceed $k_a \d^2/2$. When  $\l_m > \d/4$, $\Delta\Delta E$ is optimized at $k_m = \infty$ and can take arbitrarily large values ($\Delta \Delta E = 2 k_m \l_m \d$). The first term in \eqref{1dtheory} is the contribution from being able to actuate the (potentially) soft allosteric mode and the second term is the contribution from the two-state switch. 

\subsection{1d model at finite temperature}

In the canonical ensemble, the partition function $Z$ of a system with energy $U(x) = kx^2/2$ at temperature $T$ is given by
\beq
Z = \int e^{ -E(x)/T}dx = \int_{-\infty}^{\infty} e^{-kx^2/2T} dx = \sqrt{\frac{2\pi T}{k}},
\eeq
and its free energy is 
\beq
F = - T \ln{Z} = \frac{T}{2}\ln{\frac{k}{2\pi T}}.
\eeq
The energy of the 1d model (\eqref{eq:model}) takes the form of two parabolas that abut at $x=0$. The partition function is 
\beq
Z = \int_{-\infty}^0 e^{-V(x, -\l_m)/T} dx + \int_{0}^\infty e^{-V(x, \l_m)/T} dx
\eeq
Setting $\kappa = 2k_a + k_m$ and $\lambda(\l_m) = (k_a(\la + \lb) + k_m \l_m)/(2k_a+k_m)$ gives
\bea
Z &=& e^{-V_{\rm{min}}(-\l_m)/T} \int_{-\infty}^0 e^{-\kappa (x-\lambda(-\l_m))^2/2T} dx + e^{-V_{\rm{min}}(\l_m)/T}\int_{0}^\infty e^{-\kappa (x-\lambda(\l_m))^2/2T} dx\\
&=& \frac{1}{2} \sqrt{\frac{\pi T}{\kappa}}\Bigg[ e^{-V_{\rm{min}}(-\l_m)/T}  \left(1+\rm{erf}\left(-\sqrt{\frac{\kappa}{2T}}\lambda(-\l_m)\right)\right)+ e^{-V_{\rm{min}}(\l_m)/T} \left(1+\rm{erf}\left(\sqrt{\frac{\kappa}{2T}}\lambda(\l_m)\right)\right) \Bigg]
\eea
where $\mathrm{erf}$ is the error function.\\

In the limit where $k_m / k_a \rightarrow \infty$ the energy simplifies to 
\begin{align}
U(x) = \left\{ \begin{array}{cc} 
                \frac{k_a}{2} \left( (x - \la)^2  + (\l_m - \lb)^2 \right) & \hspace{5mm} x=-\l_m \\
                \frac{k_a}{2} \left( (x - \la)^2  + (\l_m - \lb)^2 \right) & \hspace{5mm} x=\l_m \\
                \infty & \hspace{5mm} {\rm otherwise} \\
                \end{array} \right.
\end{align}
and the free energy becomes
\beq
F(\la, \lb) = -T \ln{\left[ e^{-\frac{k_a}{2T} ((\l_m+\la)^2 + (\l_m+\lb )^2)}  + e^{-\frac{k_a}{2T} ((\l_m-\la)^2 + (\l_m-\lb )^2)}\right]}
\eeq
Taking $\l_0 = -\d/2$ and $\l_1 = \d/2$ the cooperativity is 
\beq
\Delta\Delta F = F(\d/2,-\d/2)+F(-\d/2,\d/2)-F(\d/2,\d/2)-F(-\d/2,-\d/2),
\eeq
and after some algebra,
\beq
\Delta\Delta F = 2T \ln{ \cosh{(k_a \l_m \d / T} )} .
\eeq

In the opposite limit where $k_m/k_a \to 0$,
\begin{align}
    U(x) &= \frac{k_a}{2} ( x - \la)^2 + \frac{k_a}{2} ( x - \lb)^2  \\
    &= k_a \left(x - \frac{\la+\lb}{2}\right)^2 + \frac{k_a}{4} \left(\la - \lb \right)^2\\
\end{align}
and the free energy is
\beq
F(\la, \lb) = \frac{k_a}{4}(\la - \lb)^2 - \frac{T}{2} \ln{\frac{\pi T}{k_a}}
\eeq
Taking $\l_0 = -\d/2$ and $\l_0 = \d/2$ the cooperativity is 
\beq
    \Delta\Delta F = F(\d/2,-\d/2)+F(-\d/2,\d/2)-F(\d/2,\d/2)-F(-\d/2,-\d/2)
\eeq
and after some algebra
\beq
    \Delta\Delta F = \frac{1}{2}k_a\d^2.
\eeq
When $T>0$ the transition point between mechanisms occurs for
\beq \label{ellstarVsT}
\l_m^* (T) = \frac{T}{k_a \d} \ln{\left[ e^{k_a \d^2 / 4T} + \sqrt{ e^{k_a \d^2 / 2T} -1}  \right]} 
\eeq

\subsection{Thermodynamical models}

Our models can be coarse-grained to obtain a thermodynamical description comparable to the formulation of the MWC and KNF models. This is achieved by defining ``states'' with given free energies, from which equilibrium probabilities are computed from Boltzmann's law. A state corresponds in our models to a distinct local minimum of the energy or free energy landscape. At finite temperature, a further requirement is that the free energy barriers are large relative to the temperature so that a local equilibrium occurs within the states. If two states have free energies $F_1$ and $F_2$ with a maximum $F^\dagger$ along the conformational coordinate that connects them, this assumption amounts to $F^\dagger-\max(F_1,F_2)\ll T$ in units where the Boltzmann constant is $k_B=1$. A description can also be given in terms of kinetic rates $k_{1\to 2}$ and $k_{2\to 1}$ for the transition between the states, which may for instance be of the form $k_{1\to 2}=k_0e^{-(F^\dagger-F_1)/T}$ and $k_{2\to 1}=k_0e^{-(F^\dagger-F_2)/T}$.\\

One possible coarse-graining of the 1d model could assert that two conformational states exist: state $T$ corresponding to $x<0$ and state $R$ corresponding to $x>0$. This coarse-groaning separates the two states by the barrier at $x=0$. With two binding sites, there would be a total of eight states: T00, T10, T01, T11, R00, R10, R01, R11 (the last two digits represent the state of ligand binding as in Figure \ref{theModel}). Taking $\l_0$ and $\l_1$ to be the rest lengths of the springs representing the unbound and bound states, the free energies of the eight states are,

\beq
F_{Tij} = -T \ln \left(  \int_{-\infty}^{0}  e^{- \frac{1}{2} \left(k_a (x-\l_i)^2 +  k_a (x-\l_j)^2 + k_m (x+\l_m)^2\right)/T} dx  \right)
\eeq

\beq
F_{Rij} = -T \ln \left(  \int_0^{\infty}  e^{- \frac{1}{2} \left(k_a (x-\l_i)^2 +  k_a (x-\l_j)^2 + k_m (x-\l_m)^2\right)/T} dx  \right)
\eeq

A quantity of particular interest is the fraction of occupied binding sites $Y$ at a ligand concentration $c$,

\begin{align}
Y(c) &= 0 * P_{00}(c) + \frac{1}{2} * P_{01}(c) + \frac{1}{2} * P_{10}(c) + 1 * P_{11}(c)\\
Y(c) &= \frac{ce^{-F_{T10}/T}+ ce^{-F_{R10}/T} + c^2 e^{-F_{T11}/T}+ c^2 e^{-F_{R11}/T}}{e^{-F_{T00}/T}+ e^{-F_{R00}/T} + 2ce^{-F_{T10}/T}+ 2ce^{-F_{R10}/T} + c^2 e^{-F_{T11}/T}+ c^2 e^{-F_{R11}/T}}
\end{align}

\section{Supplemental Figures}

\begin{figure}
\centering
\includegraphics[width=0.8\textwidth]{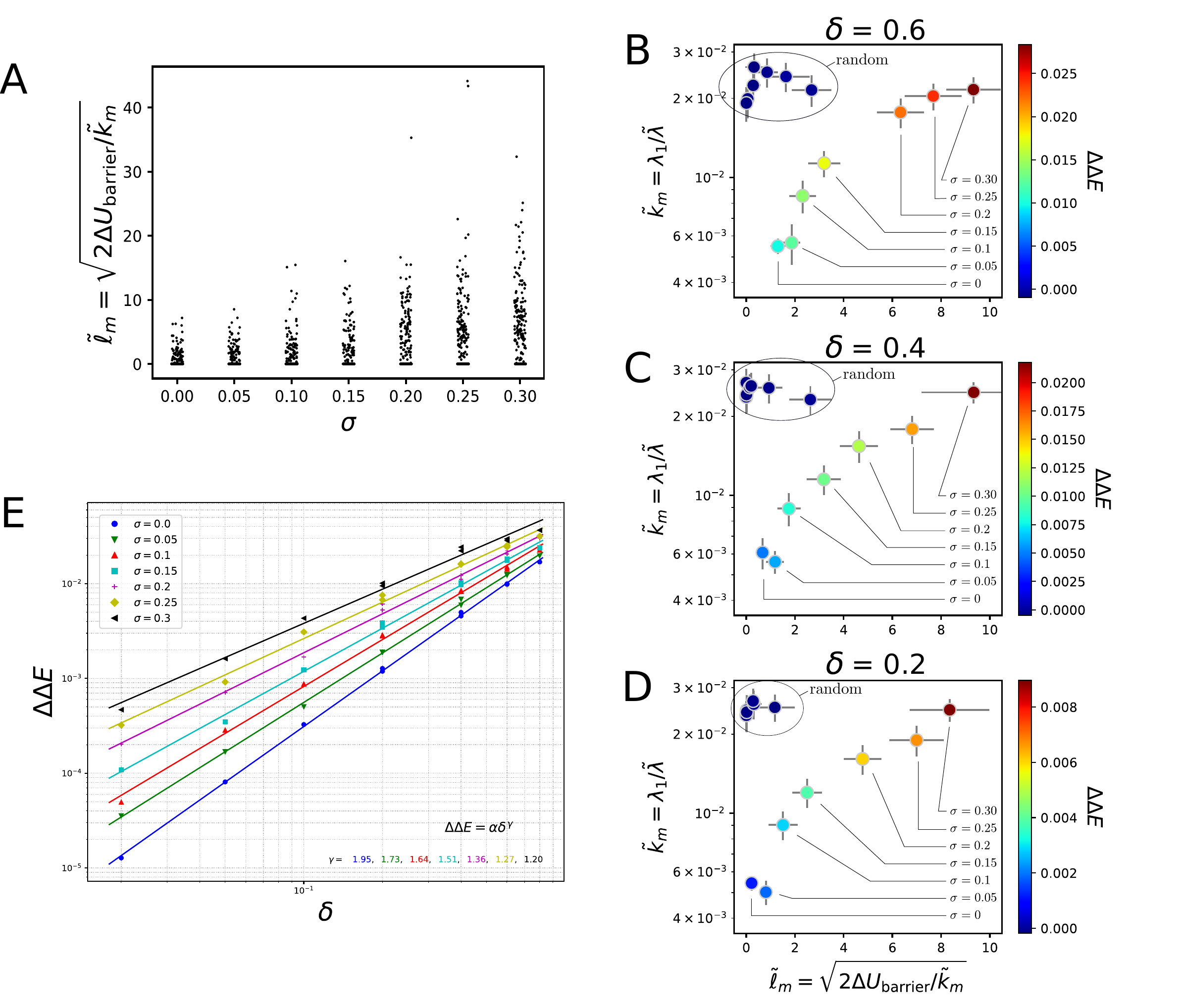}
\caption{The properties of the 1d model recapitulate the properties of the 2d elastic network.  $\Tilde{k}_m$ and $\Tilde{\l}_m$ are the 2d elastic network analogs of the 1d model's parameters $k_m$ and $\l_m$. See Methods for their definitions and motivation. (A) For each $\s$, $\Tilde{\l}_m$ of each of the 100 random (initial) and 100 evolved networks is plotted, showing an estimate of the range of possible values of $\Tilde{\l}_m$ given $\s$. (BCD) The mean $\Tilde{k}_m$ and $\Tilde{\l}_m$ and $\Delta\Delta E$ are plotted for 100 random and 100 evolved networks at different $\s$. Consistent with the 1d model, non-allosteric random networks have a large $\Tilde{k}_m$ and small $\Tilde{\l}_m$. When $\sigma$ is small, networks approach the single-state mechanism limit ($\l_m = 0, k_m=0$). As $\s$ increases, networks localize towards the multi-state mechanism limit of large $k_m$, large $\l_m$. Error bars are 95\% CI. (D) Scaling of $\Delta \Delta E$ with $\delta$. Each data point shows average an $\Delta\Delta E$ over 100 evolved networks with parameters $\s$ and $\delta$. For each $\s$, data is fit to the power law $\Delta\Delta E = \alpha \delta ^ \gamma$. The 1d model predicts that in the single-state limit, $\gamma = 2$, while in the multi-state limit $\gamma = 1$.}
\label{validation1dModel}
\end{figure}

\begin{figure}
\centering
\includegraphics[width=.8\textwidth]{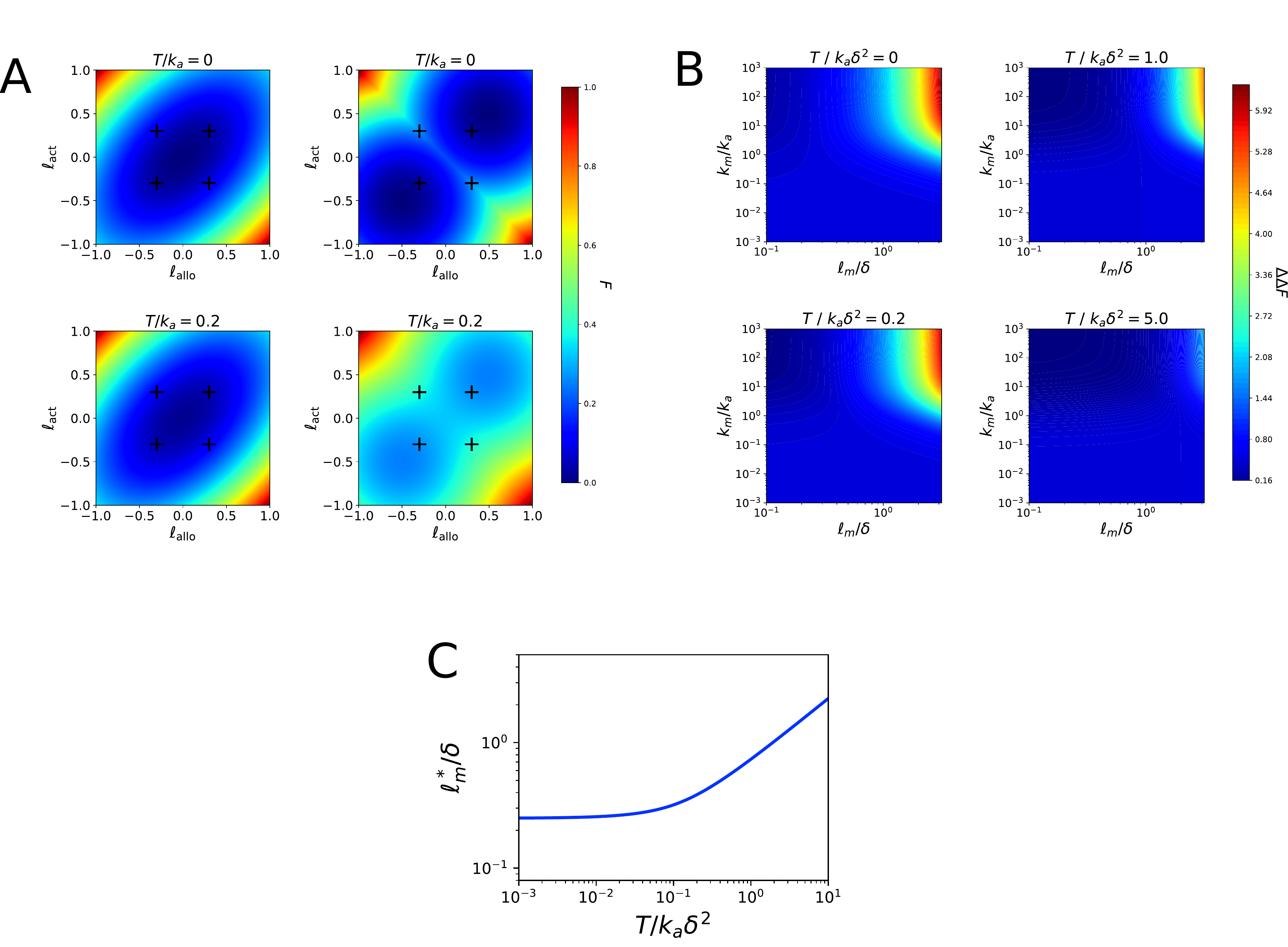}
\caption{1d model at finite temperature. (A) Free energy surfaces as a function of different binding ligands, corresponding to varying the rest lengths $\la$ and $\lb$ of the springs defining the active and allosteric sites. The two panels on the left show the free energy surfaces of a single-state mechanism ($k_m=1$ and $\l_m=0$) at $T/k_a=0$ and $T/k_a=0.2$. The two panels on the right show the free energy surfaces of a multi-state mechanism ($k_m=\infty$ and $\l_m=0.5$) at $T/k_a=0$ and $T/k_a=0.2$. (B) The cooperativity $\Delta \Delta F$ as a function of the normalized quantities $k_m/k_a$ and $\l_m / \delta$ for different temperatures. (C) A plot of $\l^*_m(T)$, the value of $\l_m$ where the single and multi-state mechanisms provide equal cooperativity at temperature $T$.}
\label{temperature}
\end{figure}

\end{document}